\documentclass[12pt,preprint]{aastex}

\newcommand{\gtsim}{\ {\raise-0.5ex\hbox{$\buildrel>\over\sim$}}\ }
\newcommand{\ltsim}{\ {\raise-0.5ex\hbox{$\buildrel<\over\sim$}}\ }

\shorttitle{ Millimagnitude Photometry for OGLE-TR-109}
\shortauthors{Fern\'andez et al.}
\begin{document}

\title{ Millimagnitude Optical Photometry for the Transiting Planetary
Candidate OGLE-TR-109
\footnote{Based on observations collected with the 
Very Large Telescope at Paranal Observatory (ESO Programme
075.C-0427(A), JMF and DM visiting observers).
}}

\author{
 Jos\'e Miguel Fern\'andez $^{1}$,
 Dante Minniti $^{1}$,
 Grzegorz Pietrzynski$^{2, 4}$,
 Wolfgang Gieren $^{2}$,
 Mar\'ia Teresa Ru\'iz $^{3}$,
 Manuela Zoccali $^{1}$, 
 Andrzej Udalski $^{4}$ and, 
 Thomas  Szeifert$^{5}$
}

\altaffiltext{1}{Department of Astronomy, Pontificia Universidad Cat\'olica, 
Casilla 306, Santiago 22, Chile\\
E-mail:  dante@astro.puc.cl, jfernand@astro.puc.cl, 
mzoccali@astro.puc.cl}

\altaffiltext{2}{Department of Physics, Universidad de Concepci\'on, Casilla 160-C, Concepci\'on, Chile
\\
E-mail: pietrzyn@hubble.cfm.udec.cl, wgieren@astro-udec.cl}

\altaffiltext{3}{Department of Astronomy, Universidad de Chile, Santiago, Chile
\\
E-mail: mtruiz@das.uchile.cl}

\altaffiltext{4}{Warsaw University Observatory, Al. Ujazdowskie 4, 00-478 Waszawa, Poland
\\
E-mail: udalski@astrouw.edu.pl}

\altaffiltext{5}{European Southern Observatory, Vitacura, Santiago, Chile
\\
E-mail: tszeifer@eso.org}

\begin{abstract}

We present precise $V$-band photometry for the low-amplitude transit
candidate star OGLE-TR-109.  This is an extreme case among the
transiting candidates found by the OGLE group because of the early
spectral type of the star (F0V), of the low transit amplitude
($A_I=0.008$ mag), and of the very short period ($P=0.58909$ days) of
the orbiting companion.  Using difference image photometry, we are
able to achieve millimagnitude errors in the individual data points.
One transit of this star is well defined in our light curve.  This
confirms the OGLE detection and rules out the possibility of a false
positive.  The measurement of this transit allows to refine the
transit amplitude ($A_V=0.006\pm 0.001$ mag), and the ephemerides for
this interesting system, as well as the radius of the possible
orbiting companion ($R_P=0.90 \pm 0.09 ~R_J$), and the inclination of
the orbit ($i=77 \pm 5$ deg).  Two other transits observed at lower
S/N confirm the period of this system measured by OGLE.  There is no
evidence for a blend of the F-type main sequence star with a redder
eclipsing binary, or for secondary transits in the present
observations.  The absence of ellipsoidal modulation in the light
curve of the primary rules out a low mass star companion or brown
dwarf with $M>14 \pm 8~ M_J$. The remaining possibilities for
OGLE-TR-109 are a blend between the F-type star and a binary with a
bluer primary star, or a new transiting extrasolar planet.

\end{abstract}

\keywords{Stars: individual (OGLE-TR-109) -- Extrasolar planets: formation}

\section{Introduction}

Udalski et al. (2002) discovered very low amplitude transits in the
$V=15.80$, $I=14.99$ magnitude star OGLE-TR-109, located in the Carina
region of the Milky Way disk, at $RA(2000)=10:53:40.73$,
$DEC(2000)=-61:25:14.8$.  They monitored 24 transits, measuring an
amplitude $A_I=0.008$ mag, and a period $P=0.58909$ days.

Based on near-infrared photometry and low-dispersion spectroscopy,
Gallardo et al. (2005) identify OGLE-TR-109 as an early F-type main
sequence star located at a distance $d=2.59\pm 0.25$ kpc, with reddening 
$E(B-V) = 0.38\pm 0.02$.  This star is intrinsically 2 magnitudes brighter 
than the Sun, with $M_V=2.55$, and also larger, with $R_s=1.5~ R_{\odot}$. 

OGLE-TR-109 has been target of a few studies due to its remarkable
location in the parameter space. It shows a very small transit amplitude
as well as very short orbital period, as illustrated in
Figure 1. Furthermore the spectral type of
the primary is the earliest so far reported for OGLE transits.

Udalski et al. (2002) estimated $R_s=1.23 ~R_{\odot}$ for OGLE-TR-109, 
and a lower limit of $R_p=0.099 ~R_{\odot}$ for its companion,
arguing that the small $R_p$ gives a larger probability
that this is an extrasolar planet. 
Upon examining in detail a number of OGLE candidates, Gallardo et al. (2005),
and Silva \& Cruz (2005) also considered the star OGLE-TR-109
as a prime planetary candidate.
Pont et al. (2005), however, classified it as an unsolved case, pointing
out three possibilities: a blend of an F-type star with a background binary,
a true planetary transit, or a false positive transit detection. 
They did not detect radial velocity changes at $\sigma=1.5$ km/s level, 
corresponding to an upper companion mass limit of $45 M_J$. They
argued that the false positive detection is the most likely explanation.
In this paper we discard this possibility, because we detect the
individual transits beyond doubt.

Estimating the expected binary star ellipsoidal modulations due to
tidal effects, Drake (2003) was the first to point out that it would
be possible to limit the mass of the secondary using accurate
photometry alone, procedure that was further refined by Sirko \&
Paczynski (2003).  In this paper we use the absence of modulation to
show that OGLE-TR-109 (if it is not a blend) should have a very low
mass companion

Section 2 presents the observations and photometry of OGLE-TR-109.
Section 3 explores the likelihood of a planetary companion.
Section 4 examines the possibility of a blend with an eclipsing binary.
Finally, section 5 gives our main conclusions.

\section{Observations and Data Reduction }

Moutou et al. (2005) demonstrated the high quality photometry that can
be obtained with the VLT. They achieved millimagnitude photometry
during the transit of OGLE-TR-132, refining its amplitude and the
planetary parameters. We are carrying out a similar program for
several recently discovered OGLE transit candidates, with OGLE-TR-109
being located in one of the four monitored fields.

The photometric observations were taken with VIMOS at the Unit
Telescope 4 (UT4) of the European Southern Observatory Very Large
Telescope (ESO VLT) at Paranal Observatory during the nights of April
9 to 12, 2005.  All four nights were clear throughout, with
sub-arcsecond seeing during most of the time. The VIMOS pixel scale is
0.205 arcsec/pixel, and no images were under-sampled during this run.
Given the large VIMOS field of view that consists of four 7$\times$8
arcmin fields covered by the four CCDs arranged in a square pattern
with a separation gap of 2 arcmin, we monitored a number of OGLE
transit candidates simultaneously.  Here we report on the observations
of OGLE-TR-109, the other stars will be discussed elsewhere.  No
standard stars were observed, because we will perform difference
photometry.

Typically 150 points per night were obtained in the field of
OGLE-TR-109, resulting in well sampled transits. The observations
lasted for about 9 hours per night, until the field went below 3
air-masses.  Figures 2 and 3 show examples of our best images taken
near the zenith, and worst images taken at very high airmass.

We used the Bessell $V$ filter of VIMOS, with $\lambda_0=5460{\rm
\AA}$, $FWHM=890{\rm \AA}$.  We choose the $V$-band in order to
complement the OGLE light curves which are made with the $I$-band
filter.  One of the main objectives of this work was to discard blends
and binary stars present among the transit candidates. For this, light
curves measured in the $V$-band can be compared with the OGLE light
curves in the I-band, and non-planetary eclipses can be discarded when
very different amplitudes are measured, for example. In addition,
while the $I$-band filter is more efficient for transit searches
(Peeper \& Gaudi 2005), the $V$-band shows better the effects of limb
darkening during the transit, and is adequate for the modeling of the
transit parameters.

In order to reduce the analysis time of the vast dataset acquired with
VIMOS, we decided to cut postage stamp images around the transit
candidates rather than process the whole images.  The future
processing of the whole images is nevertheless a gold mine to identify
additional transit candidates $\sim$2 magnitudes fainter than the OGLE
search (Fernandez et al. 2006, in preparation). We estimate that the
whole images contain $>10000$ stars with $15<V<19$ for which light
curves can be obtained with individual photometric errors $<0.01$ mag.

The images of OGLE-TR-109 analyzed here are 400$\times$400 pix, or 80
arcsec on a side.  Each of these small images contains about 500 stars
with $15<V<24$ that can be used in the difference images, and light
curve analysis.  The 7 best seeing images were selected, and a master
image was made for each night. These master images serve as reference
for the difference image analysis (see Alard 2000, Alard \& Lupton
1998).

The photometry of OGLE-TR-109, with mean $V=15.80$ gives $rms=0.002$
to $0.010$ mag throughout most of the whole run. Figure 5 shows the
full light curve for all four nights, including all data points, which
are 650 in total.  Indicated are the expected mean times of transits
as computed by Udalski et al. (2002). The large scatter at the end of
the nights is due to high airmass, and degraded image quality. There
the FWHM of the PSF grows to more than twice its original size of
about 3 pixels (0.6 arcsec). These regions of the light curve will be
excluded from the following discussion, although we note that there is
evidence for the transit at the end of night 4, the scatter is too
large to measure its duration or amplitude.

Figure 6 shows the light curve along the third night of observations,
when the best transit was monitored. Figure 7 shows the phased light
curve of the OGLE $I$-band photometry for comparison.  Although not
devoid of scatter, there are $N_t = 32$ points in our single transit
shown in Figure 6, and the reality of this transit is beyond doubt.
For a given photometric precision of a single measurement of
$\sigma_p$ and a transit depth $A$, Gaudi (2005) gives the following
equation to estimate the signal-to-noise:

$$S/N=N_{t}^{1/2} A /\sigma_p.$$ 

For OGLE-TR-109 we find the S/N of this transit to be $S/N=17$ using
$A=0.006$ and $\sigma_p=0.002$.  The $S/N$ of the transits during the
first and fourth night are poorer ($\sim 5$) merely due to the larger
scatter in the individual photometric points.

The full OGLE photometric dataset for OGLE-TR-109, covering now almost
2000 cycles (40 individual transits) make it possible to significantly
refine the ephemerides of this system:

$$HJD(middle ~of ~transit) = 2452322.55993 + 0.589128 E$$

Then, Figures 6 and 7 clearly lead to our first result: that
OGLE-TR-109 is not a false positive transit detection.  Having ruled
out that possibility of a false positive, we consider the remaining
three possibilities: a low-mass stellar companion, a planetary
transit and a blend with an eclipsing binary star.

\section{A Brown Dwarf?}

Early F-type main sequence stars are potentially interesting for astrobiology
because their habitable zones are much larger than those of later type stars
like our Sun  (Kasting, Whitmire \& Reynolds 1993).  Unfortunately,
planets have not been discovered in early type F-stars.
Planetary transits are more difficult to find (and confirm) in F-type
stars than for later type stars, because the transit amplitude decreases 
with the square of the stellar size. Thus, one expects the typical transit of
a Jovian planet to have amplitudes of the order of
$A=0.005, ~0.01$, and $0.02$ mags for F, G, and K-type 
main sequence stars, respectively. Furthermore, F-type stars have
fewer and wider lines in the optical region of the spectrum, and are
sometimes rotating very fast, which prevents from achieving the
very small radial velocity errors necessary for the detection of planets.
Therefore, it is not surprising that in the extrasolar planet lists 
there is a lack of F-type stars with respect to G and K-type stars.
There are twenty F-type stars in the Extrasolar Planets Encyclopedia
(Schneider 2005), but none earlier than F7V. Nothing is known about
the existence of planets in earlier type stars. Because they are difficult
to find, it is important to discover them.
In their favor, we should mention that usually F-type stars are 
intrinsically 1-2 magnitudes brighter that G and K-type dwarfs, and that
their larger size allows a wider range of inclination angles for observable 
transits.  

The transit time measured by Udalski et al.  (2002), shown in their
fit to the phased light curve, lasts $t_T=2.1$ hours, which is at odds
with the interpretation of a planet transiting in front of a Solar
type star($\sim 1.5$ hours for $P=0.589$). This long transit time argues
for an over sized star, either a subgiant or an earlier type main
sequence star. Gallardo et al. (2005) rules out a subgiant, on the
basis of optical and near-infrared photometry and low dispersion
spectroscopy. They favor an early F-type main sequence star with
$R_s=1.52 ~R_{\odot}$ and $T=7580\pm 350$ K. Pont et al. (2005) obtain
similar parameters, with $R_s=1.3-1.4 ~R_{\odot}$ and $T\sim 7000$ K
based on their high dispersion, low S/N spectra. We adopt the means of
the extreme values, $R_s=1.4 ~R_{\odot}$ and $T=7300$ K, because they
have been independently measured, and there is no reason {\it a
priori} to favor one over the other.  This corresponds to an F0V star
with mass $M=1.5 ~M_{\odot}$.  Note that a star of this type has a
radiative envelope, which does not require a synchronized orbit, in
spite of the short orbital period, unless the companion turns out to
be very massive (Zahn 1977). On the other hand, circularization of the
orbit must have occurred rapidly, so we assume $e=0$ for this system.

Using the full transit observed during the third night (Figure 6), we
measure $t_T = 1.8 \pm 0.1$ hours, which is consistent with $R_s=1.4
~R_{\odot}$ for $P=0.589$ days. The measured amplitude of the transit
was $A_V=0.006\pm 0.001$ mag.

We used an image of the Sun taken at the Big Bear Solar Observatory on
February 2, 2005, as a proxy to treat limb darkening realistically to
fit the light curve (Silva \& Cruz 2005). Given the period and the
estimated mass of the host star, we were able to constrain the
semi-major axis $a$ of the orbit ($2.3$ - $2.5$ $R_s$).  This short
orbit and the star size of $R_s= 1.4 ~R_{\odot}$ also determines that
the allowed range of inclination angles is relatively large. Transits
would occur for $66<i<90$ in this system. In fact, the geometric
probability of detecting a transit in a system like this is quite
large, $\sim 27\%$.  The inclination of the system is constrained by
the duration of the transit. This inclination cannot depart too much
from edge-on because the transit is long. The best fit corresponds to
a transiting object with $R_p=0.90 \pm 0.09 ~R_J$ for our adopted
stellar size of $R=1.4 ~R_{\odot}$, and an inclination of $i= 77\pm 5$
deg. We measured the transit duration from the best fit, from first to
fourth contact.

We searched the best fit minimizing chi-square in the parameter space
($R_p,a$ and $i$). To calculate the uncertainties, we fixed the values
of two parameters and let the third to vary until the change in
chi-square showed a confidence corresponding to 3-sigma.

The fit is shown in Figure 8.  Figure 9 shows the orbit of OGLE-TR-109
and its transit configuration drawn to scale for different inclination
angles: $i= 90, ~84, ~78, ~ 72,$ and $66$ degrees from top to bottom.
The smaller inclinations are ruled out because the measured transit
duration is too long for a grazing eclipse, unless the size of the
star is much larger, which appears to be ruled out by the optical-IR
photometry and spectroscopy (Gallardo et al. 2005).

The amplitude value of $A_V=0.006\pm 0.001$ mag is smaller than the
amplitude measured by Udalski et al. (2002) for OGLE-TR-109,
$A_I=0.008$ mag.  This is at odds with the expectations, because
stellar limb darkening curves in the $V$-band should be shallower in
the edges but about $10\%$ deeper in the central parts (e.g. Claret \&
Hautschildt 2003). Then, we should have obtained $A_V=0.009$ mag in
the $V$-band.  A larger discrepancy is present between the amplitudes
of OGLE-TR-132 measured by Moutou et al. (2004), $A_R=0.0066$, and
Udalski et al. (2002), $A_I=0.011$.  Therefore, given the different
methods and the errors, we ignore these differences, although it would
be interesting to explore (with larger samples of single transits
frequently sampled with millimagnitude photometry), if these
differences are evidences of systematic effects.  In particular, the
amplitude measured in the near-infrared would be very useful to test
the significance of the differences, even though measuring $A_K$ with
millimagnitude precision would be a challenge.

The smaller amplitude compared with OGLE results, and the fact that in
$V$-band the ingress and egress phases in the light curve are not so
steep (Figure 6), make our results differ from those obtained by Silva
\& Cruz (2005) for OGLE-TR-109 ($R_p=1.18 ~R_J$ and $i=89.0$).

Table 1 summarizes the results obtained from the light curves,
giving the period, HJD, transit time, orbital
inclination, amplitude, and radius of the companion.

Table 2 lists the previous estimates of the size for the OGLE-TR-109
companion from the literature, and this work. No serious disagreement
is found among the four existing measurements of $R_p$.  In fact, all
quoted sizes are substellar, consistent with giant planets or brown
dwarfs. As for giant planets is concerned, OGLE-TR-109-b does not seem
to be over sized. One would expect that over such close orbit with
$a=0.016 ~AU$ the planet should be inflated by stellar radiation. If
this is indeed the case, we could be in the presence of a very low
mass planet.

The companion orbits at a distance of merely $a=0.016 ~AU ~=2.4 ~R_*$.
Then, the planet is orbiting just outside of the stellar Roche
radius. This represents even a more extreme case than those discussed
by Patzold, Carona, \& Rauer (2005). They examined the tidal
interaction of close-in OGLE planets, concluding that OGLE-TR-56-b,
for example, would spin-up the stellar rotation and spiral into the
star in a short timescale. It remains to be studied if a similar fate
may await OGLE-TR-109-b.

The shortest period planets known are OGLE-TR-56, and OGLE-TR-113
(Udalski et al.  2002, Konacki et al. 2003, Torres et al. 2004a, Bouchy
et al. 2004, Konacki et al.  2004). With $P=1.2~ d$, and $P=1.4~ d$,
their orbital periods more than double that of OGLE-TR-109.  If this
is a planetary transit, it would be the shortest planet known so far,
around the hottest star measured. Unfortunately, the star is rapidly
rotating, with $V=35.4\pm 1.8$ km/s, which makes the measurement of
precise velocities very difficult (Pont et al. 2005). Therefore, in
this (and other similar cases that might arise), it is rather
important to find other means of establishing the presence of a planet
that do not require the measurement of $M_p ~sin ~i$ by radial
velocities.

One such possibility is to place limits to the ellipsoidal modulation
of the light curve due to tidal effects, as discussed by Drake (2003)
and Sirko \& Paczynski (2003). A short period massive companion would
induce a detectable photometric signature in the light curve of the
primary star, with a periodicity of half the orbital period. For
OGLE-TR-109 Sirko \& Paczynski (2003) measure one of the light curves
with least modulations.  No ellipsoidal modulation with $P=0.3 ~days$
is detected in our light curve for OGLE-TR-109 larger than $\sim
0.001$ mag (Figure 10), supporting their conclusions. A caveat is that
our data are not ideal to measure low amplitude modulations of the
light curve, as it is well known that contiguous photometric points
could be correlated, the OGLE dataset is more useful in this respect.

Drake (2003) modeled the expected amplitudes of the modulation due to
ellipsoidal effects for main sequence stars of different spectral
types as function of their companion mass. The effect is larger for
early type main sequence stars, because their surface gravity is lower
than that of a solar-type star, for example.  He considered a single
period of 2 days, but the modulation increases with decreasing period
as $P^{-2}$, and one would expect for his models to provide a strong
upper limit for the case of OGLE-TR-109 with $P=0.589$ days. In fact,
the effect in the light curve amplitudes for this star should be more
than $11$ times that shown in Figure 1 of Drake (2003). For example,
an F0V star with a $0.1 ~M_{\odot}$ companion would exhibit a
modulation with amplitude of $0.4 ~mmag$ if the orbital period is $P=2
~days$. For an orbital period of $P=0.589 ~days$, this amplitude would
be about $4.5 ~mmag$, certainly not observed in the light curve of
OGLE-TR-109.  Figure 10 shows the phased light curve of OGLE-TR-109 for
nights 2, 3, and 4, along with the expected modulation corresponding
to a companion with $M=0.08 ~M_{\odot}$ (the brown dwarf limit), and
$0.045 ~M_{\odot}$ (the radial velocity limit of Pont et al. 2004).
These modulations, with amplitudes 0.0034 and 0.0023 mag,
respectively, are not seen in our light curve (Figure 10), indicating
that the companion mass is smaller than these limits.

The OGLE light curve provides an even more stringent limit to the
mass.  Figure 11 shows the full OGLE dataset binned in 0.02 phase bins
with parts around the transit removed. Typically 30--50 observations
were included in each bin. The light curve is folded with the basic
period and twice its value (as if the star was an eclipsing binary
with equal depth eclipses) -- lower and upper panels,
respectively. The thick line is an ellipsoidal variation of $1.0$ mmag
amplitude. It is clear from this plot that the data do not show such
variability even at a much lower level.  A formal fit to the raw data
indicates no significant ellipsoidal variation larger than $0.5$ mmag
for both periods. This is in good agreement with the result of Sirko
\& Paczynski (2003): $a_{c2}=0.61\pm 0.33$~mmag. Using the latter
amplitude to extrapolate in Figure 1 of Drake (2003) for an F-type
main sequence star yields an upper limit of $M=0.014\pm0.008 M_\odot$
for the companion mass.  This rules out a low mass star or brown
dwarf, and leaves the possibility of a transiting giant planet.  Note
that this constraint from the ellipsoidal modulation agrees with the
maximum mass limit $M_{max}=0.045 ~M_{\odot}$ established for the
companion by Pont et al. (2005) using radial velocities.

Interestingly, the reflection or heating effect due to the planet
might be detectable because of the short orbital axis. The parameter
$a_{c1}$ of Sirko \& Paczynski (2003) measures this effect. In the
case of OGLE-TR-109 with $a=0.016$ AU (Silva \& Cruz 2005), the planet
with $R=0.9 ~R_J$ intercepts about 1/1000 of the stellar flux, and one
would expect $a_{c1}=1$ mmag. This is not very different from
$a_{c1}=0.78\pm 0.35$ mmag measured by Sirko \& Paczynski (2003).

\section{A Blend with an Eclipsing Binary?}

Unfortunately, a blend of an F-type main sequence star with a
background binary is difficult to rule out. Mandushev et al. (2005)
clearly show how a very good transit candidate with consistent radial
velocities finally was a blended system. In the absence of radial
velocities, the possibility of a blend with a background eclipsing
binary star can be explored by detecting secondary transits, by
comparing the amplitudes of the individual transits, and by comparing
the amplitudes of a single eclipse measured with two different
filters. 

Alternated eclipses of different depths (i.e.  presence of primary and
secondary eclipses), would be evidence of an eclipsing binary with
period $P=2\times0.59=1.18$ days. Note that the eclipses in the first
and fourth night do not have high enough quality to compare their
amplitudes with the transit shown in Figure 6.  However, the OGLE data
do not indicate different depth of eclipses within photometric errors.

In addition, the best parts of our light curves in nights 1, 2 and 3
monitor the times when the secondary eclipses are expected to happen.
No secondary eclipses are detected in the light curves of night 2
(with the smallest scatter), with amplitudes larger than $\sim 0.001$
mag. No secondary eclipses are apparent in the OGLE light curves
either.

In some cases a blend with an eclipsing binary should show the heating
effects on the secondary, particularly for short period binaries with
$P\approx 0.6$ days. However, the reflection effect measured by Sirko
\& Paczynski (2003) for OGLE-TR-109 is also small, $a_{c1}=0.78$, as
mentioned above.

At first sight one argument in favor of a blend with an eclipsing
binary in the background might the difference between the amplitudes
measured in the $V$ and $I$-bands.  Udalski et al. (2002) obtained
$A_I=0.008\pm 0.001$.  Due to limb darkening, the total transit
amplitude in the $V$-band should be about $10\%$ larger than that, or
$A_V=0.009\pm 0.001$.  Comparing this value with $A_V=0.006\pm 0.001$
measured here, there is a $2.1\sigma$ difference.  This is not
significant enough to discard OGLE-TR-109 as a blend.

In order to produce the low amplitude transits, the F0V star is
located in the foreground and dominates the flux, while the blend must
be with a background main sequence star eclipsed by a late-type main
sequence star or a giant eclipsed by a main-sequence star.

If the blend is with a later type main sequence star (G or K) it would
produce $A_V$ is larger than $A_I$ (e.g. Drake 2003), contrary to the
observations. The fact that $A_V=0.006 ~mag$ is smaller than
$A_I=0.008 ~mag$ would indicate that the binary has a primary star
that is fainter and bluer than the F0V star. This scenario must be
seriusly considered, because it has been observed in other transiting
candidate, OGLE-TR-33, which is a hierarchical triple system composed
of a slightly evolved F6 star and an eclipsing binary with a K7-M0
star orbiting an F4 star (Torres et al. 2004b). Only precise radial
velocities can rule out (or confirm) this kind of configuration.

The eclipsed red giant scenario is ruled out because of the same
reason of the redder background binary , and because it would be
bright enough to see in the spectra of OGLE-TR-109.

An additional piece of evidence against a blend with a redder
background binary star is given by Gallardo et al.  (2005), who do not
detect significant near-infrared excess in the $K$-band.  They
measured $V-K_0=0.51\pm 0.1$, and $E(B-V) = 0.38\pm 0.02$, adequate
for an F0V star in the Milky Way disk.

Another test of binarity or contamination by a background source is to
measure the movement of the photocenter during transit with respect to
the uneclipsed portions of the light curve. This method, often useful
for microlensing events (Alcock et al. 2003), does not yield anything
conclusive in this regard because the OGLE-TR-109 transit amplitude is
so small.

In order to further explore the blend issue, we have acquired Ks band
images with SOFI at the ESO New Technology Telescope in May
2005. Figure 4 shows the same field of view of Figure 2 and 3 in the
Ks band. The small circle indicates a very red point source,
presumably a late M-type main sequence star that was not present in
the V-band images. This source, however, cannot be the source of the
TR109 eclipse. There are no other faint red stars within 1 arcsec that
would contaminate the photometry of our main target star.

Other sources of variability that may resemble low amplitude transits are
star spots. These should not be constant with time, however, as has been
observed with the transits of OGLE-TR-109. Furthermore, star spots should vary
with the rotation period of the star, which in this case should be
$P_{rot}\approx 2 ~days$, in contradiction with the observations.

To summarize, we have shown that for this special case, the photometry
shows no evidence for a blend of the F-type star with a background
redder binary, and provides tight enough constraints to discard a
massive companion, so OGLE-TR-109-b is most likely a blend between an
F-type star and a binary with a bluer primary star or a transiting giant
planet.

\section{Conclusions}

Udalski et al. (2002) discovered low amplitude transits in the F0 main
sequence star OGLE-TR-109, which we observed with VIMOS at the ESO
VLT.  The systematic effects dominate over photon noise because
OGLE-TR-109 is a relatively bright star. However, we are able to
clearly detect the low amplitude transits on this star.  We note that
this approach is complementary to the one followed by Udalski et
al. (2003). They observed 24 transits, but with few points per
transit. Here we observe only three transits, but with numerous points
during the transit.  Given the similar results obtained by the two
widely different approaches, it is our strongest conclusion that
OGLE-TR-109 is not a false positive transit detection as argued by
Pont et al. (2005).

Examining the transit depths and possibility of secondary eclipses, we
suggest that the presence of a blend of the F-type star with a redder
eclipsing binary is unlikely, but a blend with a binary with a bluer
primary star cannot be ruled out by now.

Drake (2003), and Sirko \& Paczynski (2003) showed that it is possible
to limit the mass of the secondary using accurate photometry alone.
Analyzing the limits to the ellipsoidal variability measured by Sirko
\& Paczynski (2003), and the simulations of Drake (2003), we conclude
that a low mass stellar companion with $M_p > 0.014 \pm 0.008
~M{\odot}$ is ruled out, in agreement with the radial velocity limit
of $M_p \leq 45 ~M_J$ measured by Pont et al. (2005).  The precise
mass of the planet may remain undecided for quite some time, because
the rapid rotation of the primary star precludes the measurement of
radial velocities accurate enough to measure the mass of the
companion.

Finally, we improve upon the parameters of the transit, in particular
measuring a transit time, $t_T=1.8\pm 0.1$ hours, that is in agreement with
the orbital and stellar parameters.  We also update the ephemerides
and orbital parameters of this interesting system, measuring $P=0.58909$
days, $A_V=0.006\pm 0.001$ mag, $HJD= 2452322.55993$, and $i=77\pm 5$ deg.

We have identified two posible scenarios for OGLE-TR-109: a F-type
main sequence star blended with a binary with a bluer primary star, or
a giant planet orbiting with a very short period around a F-type main
sequence star. This object would be unique in its class so far.  The
existence of a low mass companion orbiting an F0V star with $P=0.589$
days pushes not only the limits for planetary formation and migration
even further, but also the atmosphere of this object would be beyond
the models considered so far. The fate of this object is interesting
as well, because tidal interaction may force the accretion by its
mother star in a relatively short timescale.

\acknowledgments{
DM, JMF, GP, MZ, MTR, WG are supported by Fondap Center for Astrophysics 
No. 15010003. DM also thanks the John Simon Guggenheim Foundation.
Partial support to the OGLE project was provided by the Polish MNII
grant 2P03D02124, NSF grant AST-0204908 and NASA grant NAG5-12212. A.U.
acknowledges support from the grant ``Subsydium Profesorskie'' from the
Foundation for Polish Science.
We thank the ESO staff at Paranal Observatory.
}

\clearpage
\begin{deluxetable}{lllllllll}
\tablewidth{0pt}
\tablecaption{Measured Parameters for OGLE-TR-109-b}
\tablehead{
\multicolumn{1}{c}{Parameter}&
\multicolumn{1}{c}{Value}&
\multicolumn{1}{c}{Comment}}
\startdata
$P$ (d)              & $0.589128$        & \nl
$JDo$ (d)           & $2452322.55993\pm 0.001$& \nl
Transit duration (hr) & $1.8\pm 0.1$  & \nl
transit amplitude (mag)       & $0.006\pm 0.001$      & \nl
Inclination (deg) & $77\pm 5$             & \nl
Companion radius $R_p$ ($R_J$)             & $0.90\pm 0.09$ & \nl
Companion mass ($M_J$)            & $<45 $            & Pont et al. (2004) \nl
Companion mass ($M_J$)            & $<14 \pm 8$      & \nl
\enddata
\end {deluxetable}

\begin{deluxetable}{llll}
\tablewidth{0pt}
\tablecaption{Radius of OGLE-TR-109-b}
\tablehead{
\multicolumn{1}{c}{Reference}&
\multicolumn{1}{c}{$R_p (R_J)$}&
\multicolumn{1}{c}{Comments}}
\startdata
Udalski et al. (2002) &$0.99$         & lower limit, 24 transits, $I$-band photometry\nl
Gallardo et al. (2005)&$1.31\pm 0.10$ & using OGLE amplitude, new stellar parameters\nl
Silva \& Cruz (2005)  &$1.18$         & re-analysis of the OGLE $I$-band photometry\nl
This work             &$0.90\pm 0.09$ & single transit, $V$-band photometry\nl

\enddata
\end {deluxetable}

\clearpage
\begin{figure}
\plotone{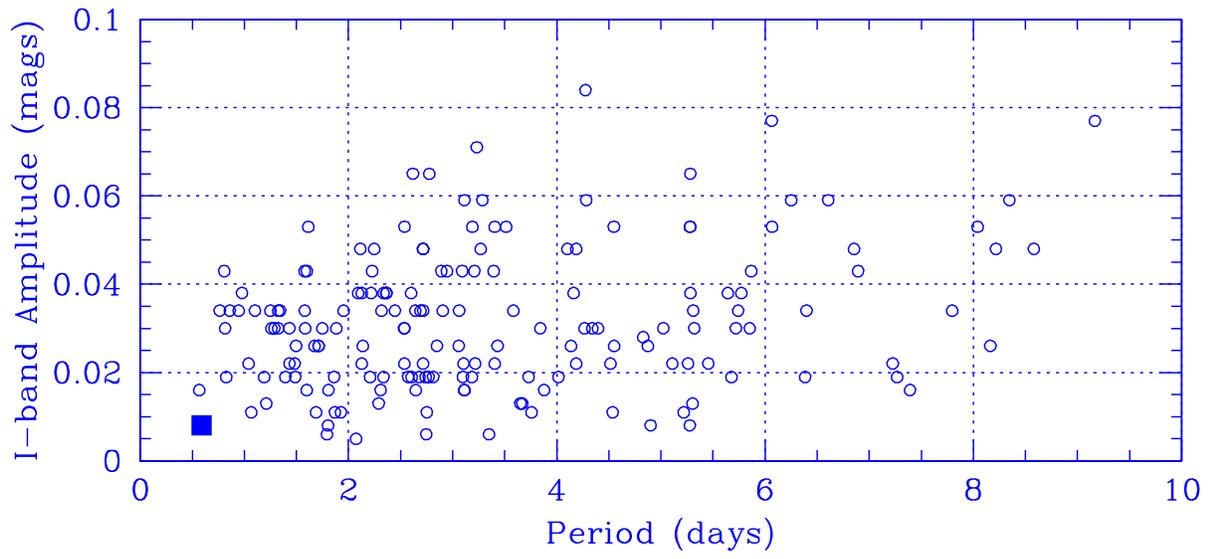}
\caption{
Amplitudes and periods of the OGLE transit candidates. The star
OGLE-TR-109 is indicated with  the solid square.
}
\end{figure}

\begin{figure}
\plotone{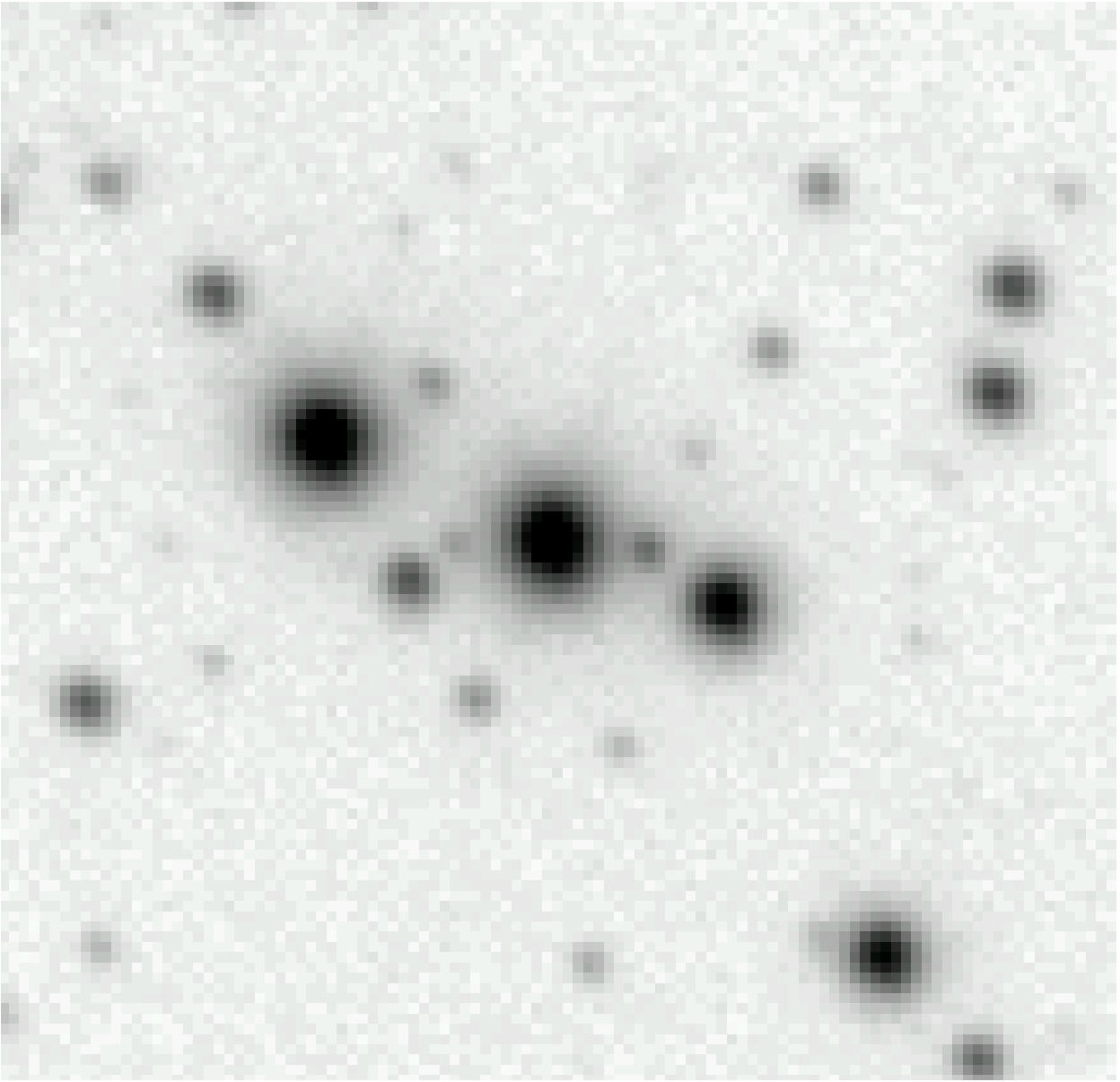}
\caption{
Finding chart obtained with VIMOS for OGLE-TR-109 ($V=15.80$),
which is the bright star at the center. This image covers
$20\times 20$ arcsec, and
represents the best FWHM obtained ($0.5$ arcsec).
The faintest stars seen in this image have $V\sim 24$.
}
\end{figure}

\begin{figure}
\plotone{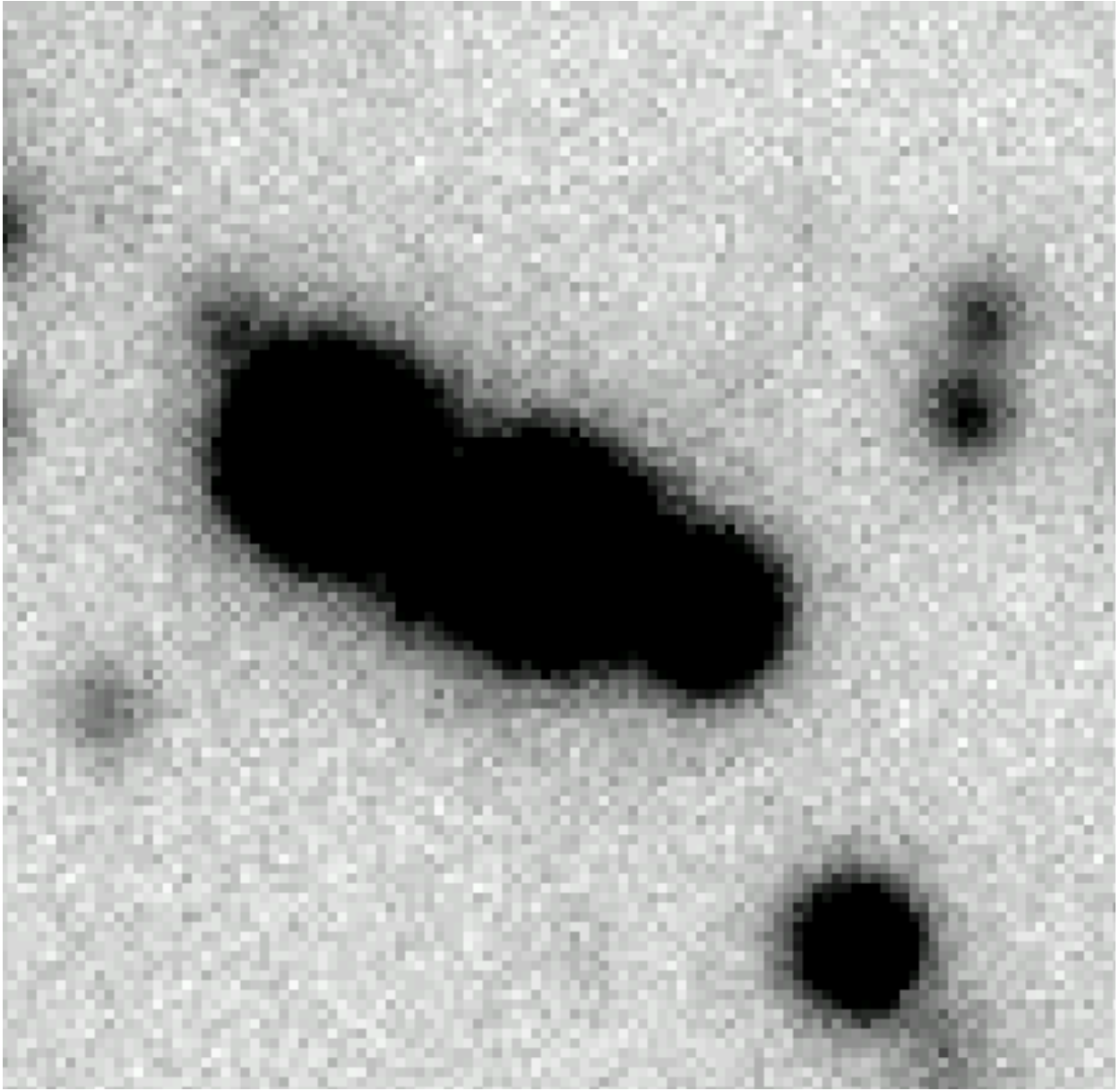}
\caption{
Same as Figure 2. This image covers $20\times 20$ arcsec, and
represents the worst FWHM obtained ($1.6$ arcsec) at high airmass ($X\sim3$).
The faintest stars seen in this image have $V\sim 20$.
There is clear blending of the flux from the two bright companion stars
to OGLE-TR-109. Even using the difference image analysis, 
these images yield very poor photometry 
for our target ($rms \sim 0.02$), rendering them useless when trying to 
measure the transits to millimagnitude accuracy.
}
\end{figure}

\begin{figure}
\plotone{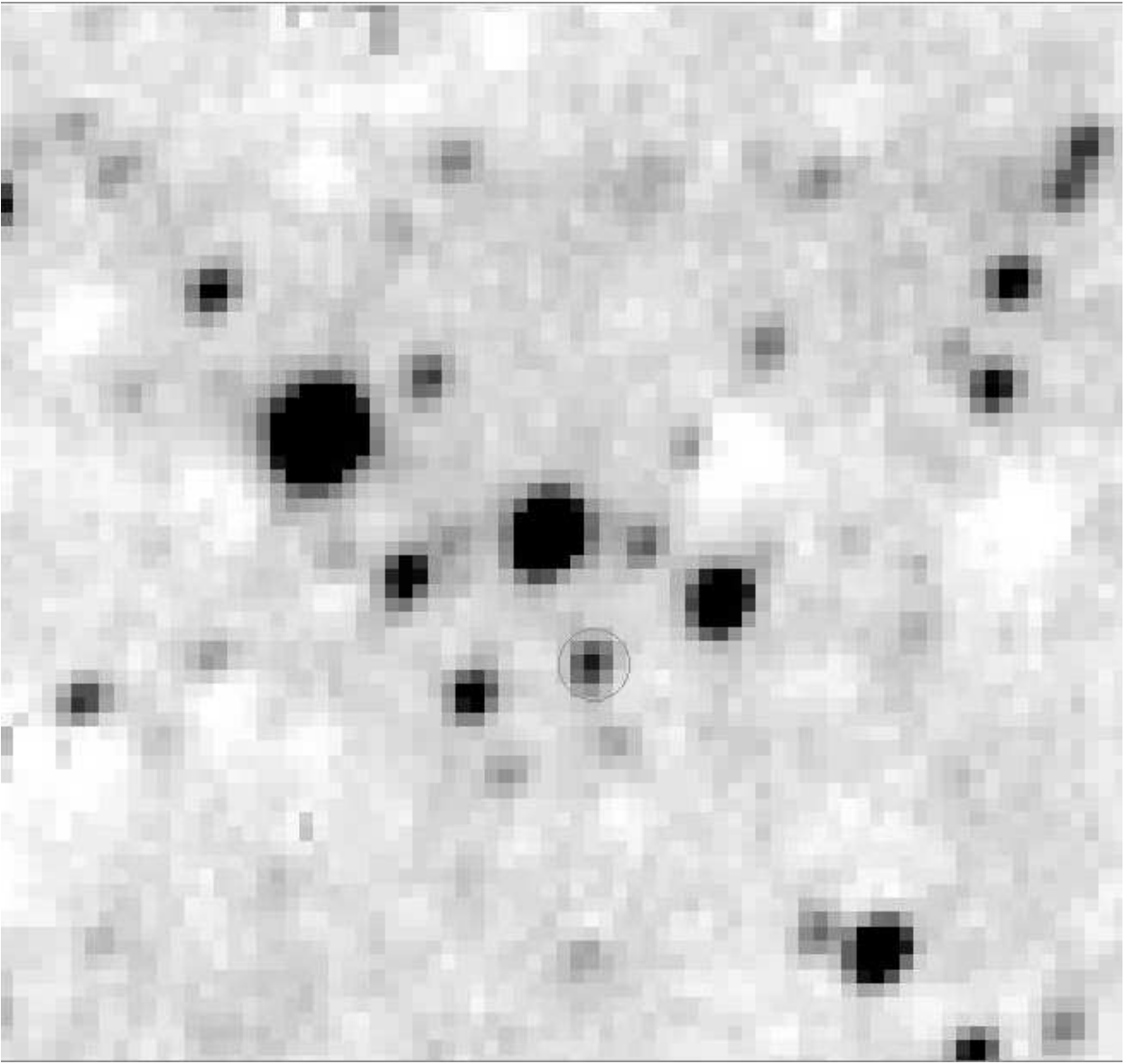}
\caption{
Same field of view of Figure 2 and 3 in the
Ks band. The small circle indicates a very red point source,
presumably a late M-type main sequence star that was not present
in the V-band images.
}
\end{figure}

\begin{figure}
\plotone{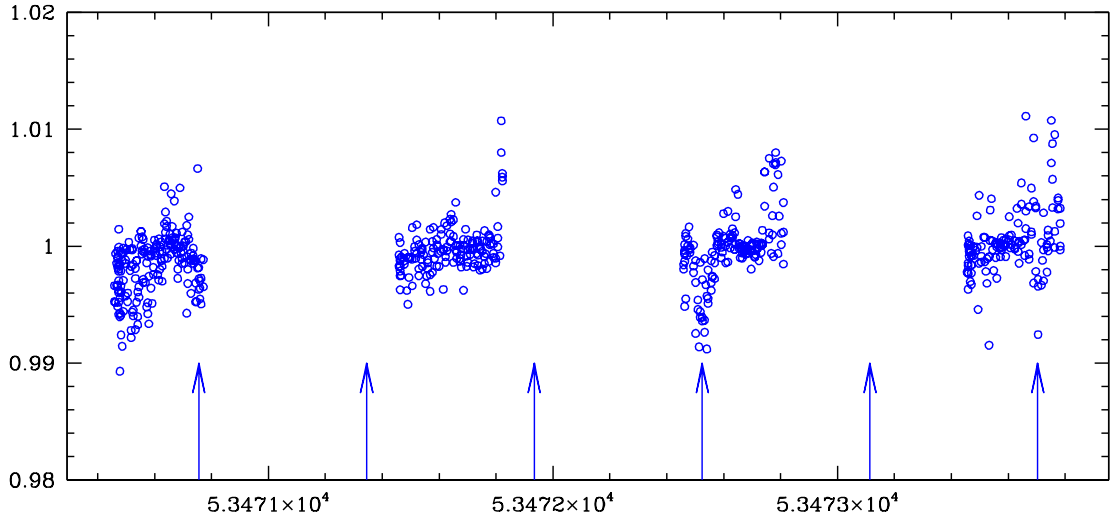}
\caption{
Light curve of OGLE-TR-109 during the four nights.
All 650 points measured are plotted, including points obtained
at high airmass at the end of the nights, for which the diminished 
quality of the photometry is obvious. 
The third night shows a well monitored transit.
The predicted mean transit times from OGLE are shown with the arrows.
}
\end{figure}

\begin{figure}
\plotone{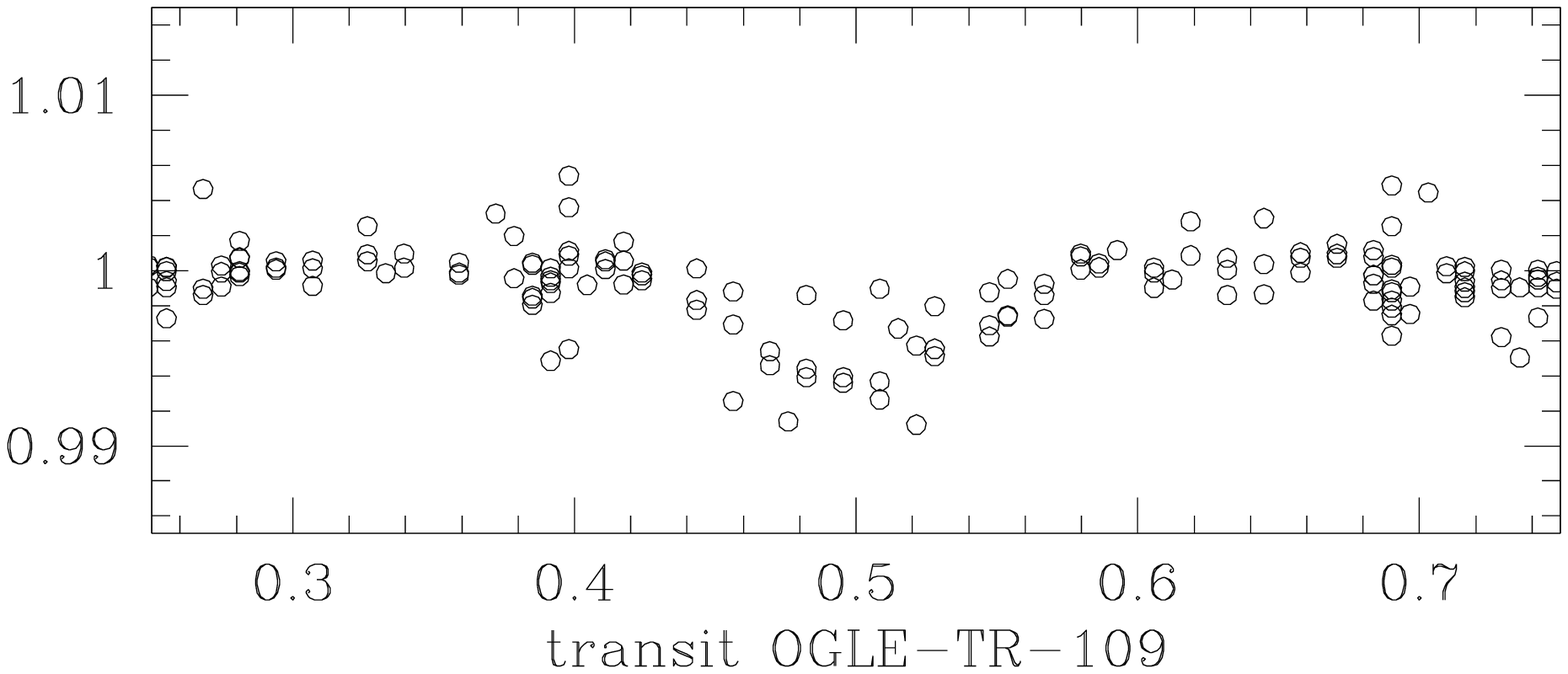}
\caption{
Best single transit of OGLE-TR-109 observed during the third night with VIMOS.
}
\end{figure}

\begin{figure}
\plotone{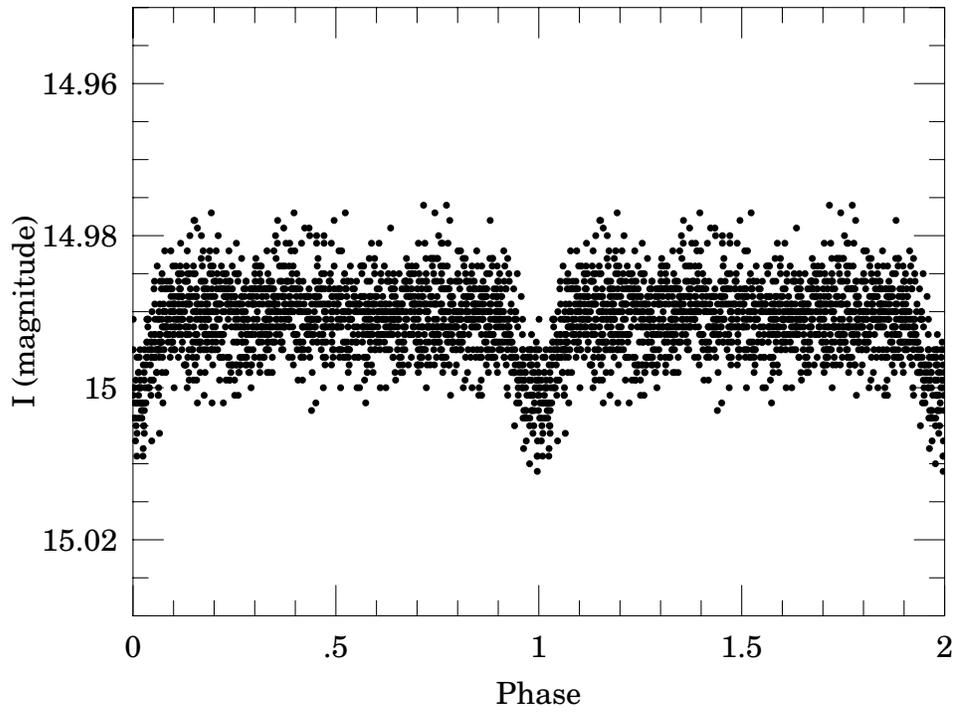}
\caption{
New phased light curve of OGLE-TR-109 for the OGLE data in the $I$-band.
}
\end{figure}

\begin{figure}
\plotone{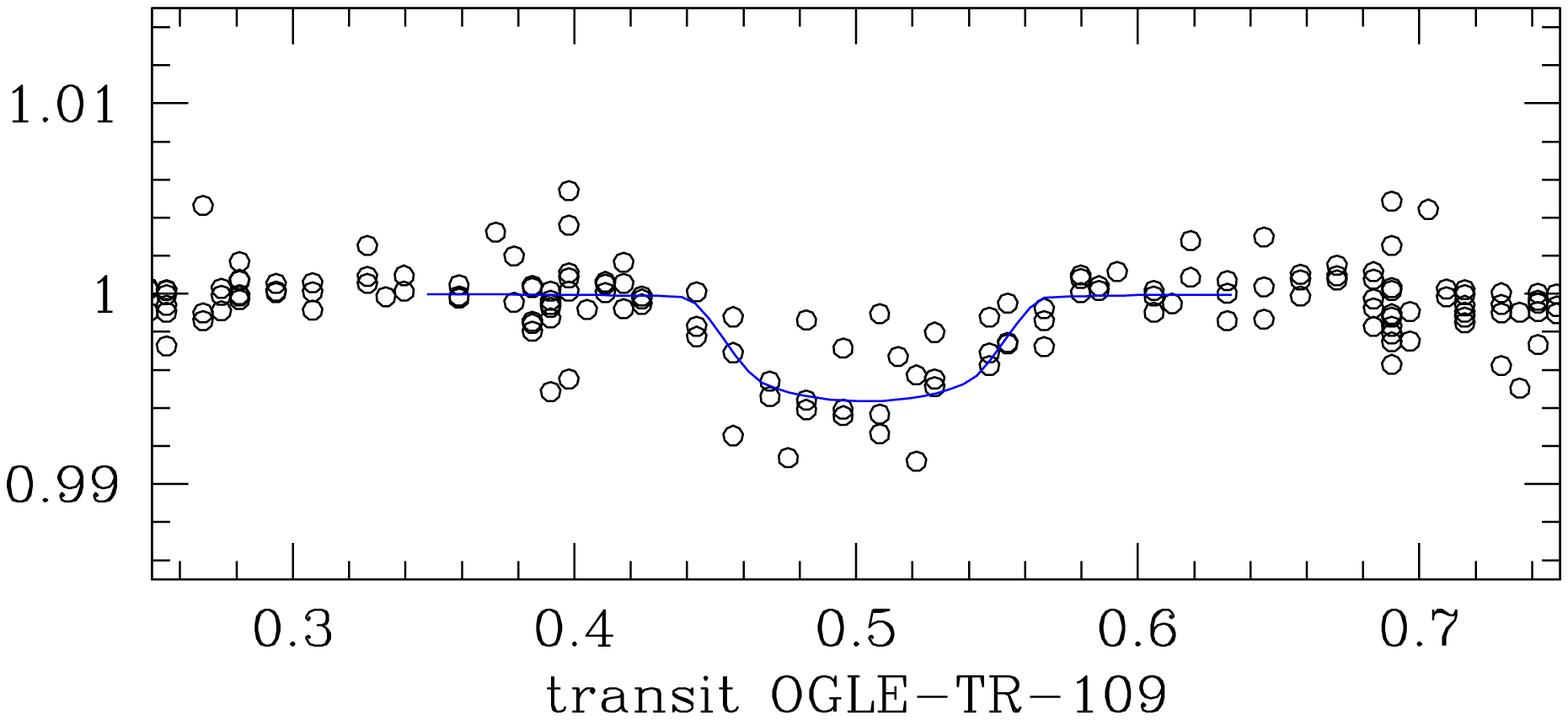}
\caption{
Fit to the best single transit of OGLE-TR-109 observed 
during the third night with VIMOS.
}
\end{figure}

\begin{figure}
\plotone{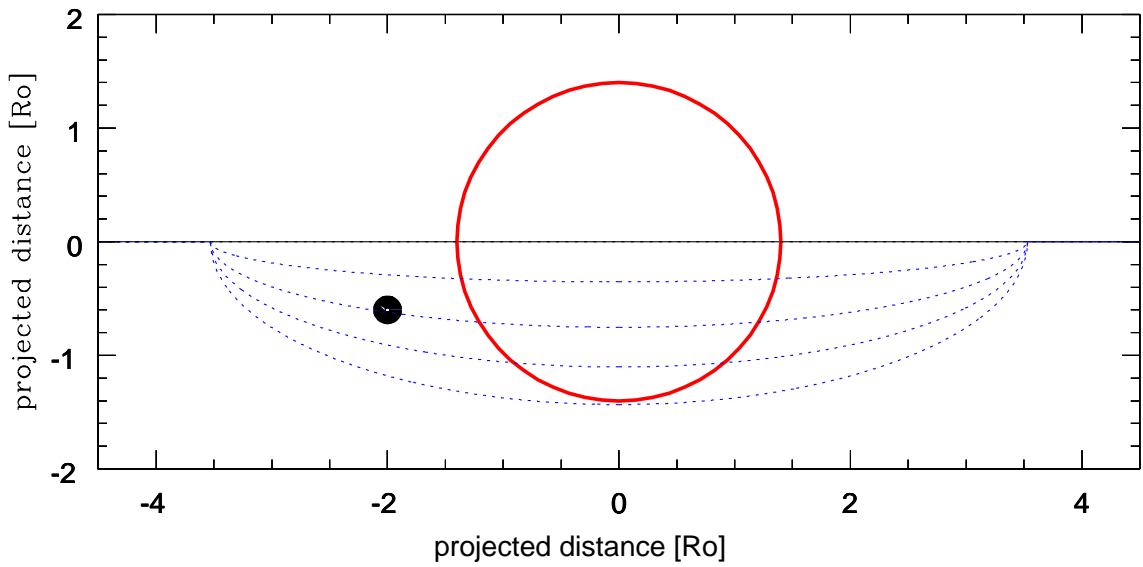}
\caption{ 
Forefront half of the OGLE-TR-109 orbit (and transit
configuration) drawn to scale for different inclination angles: $i=
90, ~84, ~78, ~ 72,$ and $66$ degrees, from top to bottom.  The
smaller inclinations are ruled out because the measured transit
duration is too long for a grazing eclipse. The larger inclinations
are ruled out because of the inclined ingress and egress phases.  Note
that the trajectory in front of the star departs from linearity
because of the short orbital axis.}
\end{figure}

\begin{figure}
\plotone{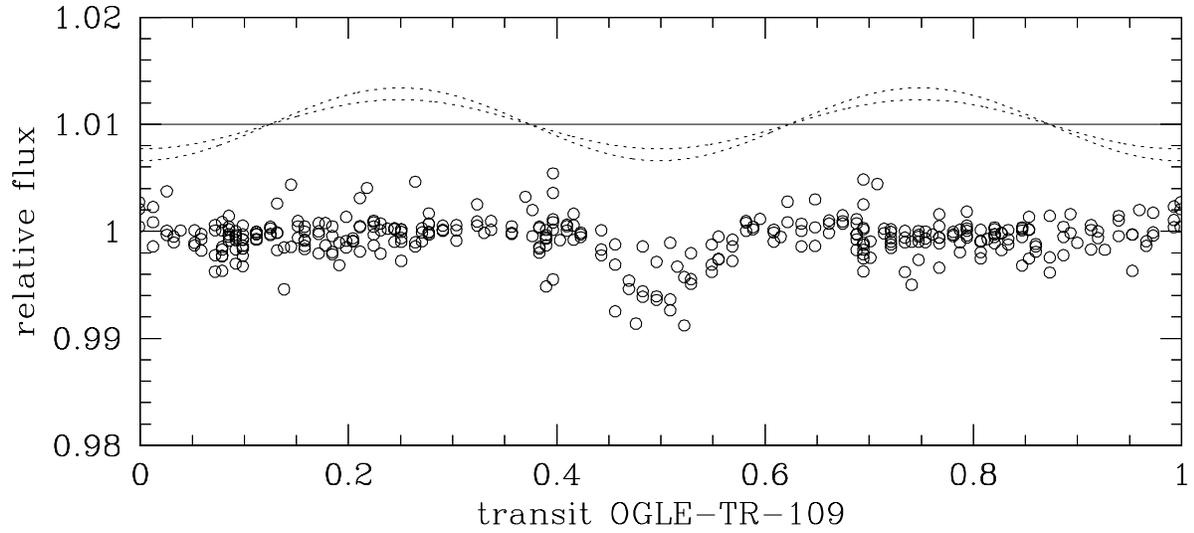}
\caption{
Phased light curve of OGLE-TR-109 for nights 2, 3, and 4.
For comparison, the expected ellipsoidal modulation corresponding to
a companion with $M=0.08$ and $0.045 ~M_{\odot}$ are shown with the
dashed lines above the light 
curve, with amplitudes 0.0034 and 0.0023 mag, respectively. The fact
that no such light modulations are seen allows to constrain the mass of
the companion.
}
\end{figure}

\begin{figure}
\plotone{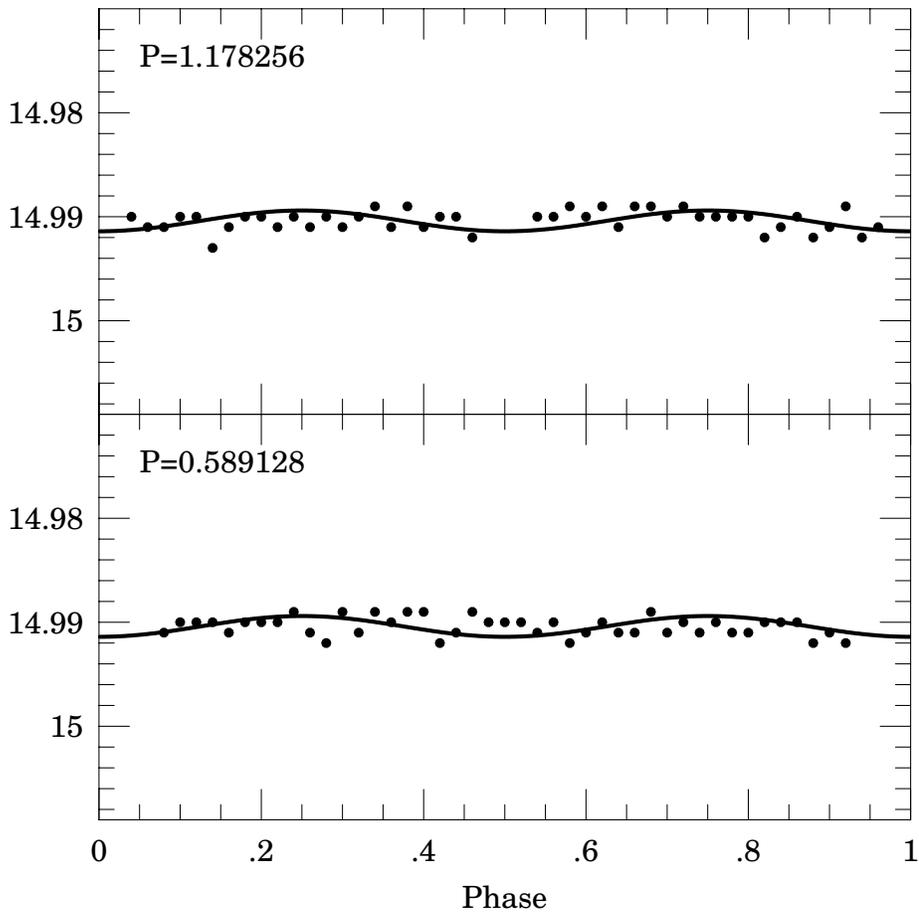}
\caption{
Light curve of OGLE-TR-109 for the OGLE data in the $I$-band,
binned in 0.02 phase bins with parts around the transit removed.
The light curve is folded with the basic period and twice
its value. The thick lines show the expected ellipsoidal modulation with
amplitude of $1.0$ mmag.
}
\end{figure}

\end{document}